\documentclass[aps,prd,groupedaddress]{revtex4}
\usepackage{amsmath}
\usepackage{amssymb}
\usepackage{bbm}
\usepackage{epsfig}
\begin{document}
\begin{flushright}
SHEP-08-38\\
\end{flushright}
\title{Dimension six FCNC operators and top production at the LHC}

\author{R.A. Coimbra$^{1,2}$~\footnote{rita@teor.fis.uc.pt}, P.M. Ferreira$^{3,4}$~\footnote{ferreira@cii.fc.ul.pt}, R.B.
Guedes$^{4}$~\footnote{renato@cii.fc.ul.pt},
 O. Oliveira$^{2}$~\footnote{orlando@teor.fis.uc.pt},
A. Onofre$^{1}$~\footnote{onofre@lipc.fis.uc.pt},
   R.
Santos$^{5}$~\footnote{rsantos@cii.fc.ul.pt} and Miguel
Won$^{1,5}$~\footnote{miguel.won@lipc.fis.uc.pt}} \affiliation{
$^{1}$  LIP- Departamento de F\'{\i}sica, Universidade de Coimbra, 3004-516 Coimbra, Portugal;\\
$^{2}$  Centro de F\'{\i}sica Computacional, Universidade de Coimbra, 3004-516 Coimbra, Portugal;\\
$^{3}$  Instituto Superior de Engenharia de Lisboa, Rua
Conselheiro Em\'{\i}dio Navarro 1, 1959-007 Lisboa, Portugal;\\
$^{4}$ Centro de F\'{\i}sica Te\'orica e Computacional, Faculdade
de Ci\^encias, Universidade de Lisboa,\\ Avenida Professor Gama
Pinto, 2, 1649-003
Lisboa, Portugal; \\
$^{5}$ NExT Institute and School of Physics and Astronomy,
University of Southampton Highfield, Southampton SO17 1BJ, United
Kingdom}

\date{\today}

\begin{abstract}
\noindent
In this work we calculate the effects of the electroweak flavour
changing neutral currents dimension six effective operators on
single top production at the LHC. These results are then combined
with previous ones we have obtained for the strong sector. This
allow us, for the first time, to perform a combined analysis of
flavour changing neutral currents in top production with all
contributing dimension six operators. Finally, we study the
feasibility of their observation and characterization both at the
Tevatron and at the LHC.
\end{abstract}
\pacs{PACS number(s): }
\maketitle

\section{Introduction}
\label{sec:intr}  CERN's Large Hadron Collider (LHC) will soon start
colliding protons with 14 TeV center of mass energy. This top quark
factory will allow us to study the least known of all quarks with
unprecedented precision. The study of flavour changing neutral
currents (FCNC) is one of the most interesting research topics
related to top quark physics as a wide variety of models shows a
strong dependence in the measurable FCNC quantities. In fact, the
top quark FCNC branching ratios can vary from extremely small in the
Standard Model (SM) to measurable values at the LHC in some of its
extensions.

 In a series of
papers~\cite{Ferreira:2005dr,Ferreira:2006xe,Ferreira:2006in,
Ferreira:2008cj} we studied flavour changing single top production
using the effective operator formalism \cite{buch}.
In~\cite{Ferreira:2005dr,Ferreira:2006xe,Ferreira:2006in} we assumed
that new physics would originate from strong interactions alone.
Consequently, we have calculated the FCNC top decay branching ratio
$t \rightarrow u (c) \, g$ as well as all cross sections for single
top production using the new dimension six operators stemming from
the strong sector. These production processes included associated
production of a single top quark alongside a jet, a Higgs boson or
an electroweak gauge boson. The main conclusion was that for large
values of BR($t \rightarrow q \, g$), with q = u, c, these processes
might be observable at the LHC.

 Not all of the processes calculated
for the LHC receive contributions from new strong interactions alone
- some processes are also modified by dimension six operators from
the electroweak sector, that is, by operators that do not include
the strong field tensor. In a recent paper~\cite{Ferreira:2008cj} we
have investigated the possibility of distinguishing the
contributions from the strong and from the electroweak sectors for
the FCNC processes $pp \rightarrow t \, Z$ and $pp \rightarrow t \,
\gamma$. We found new physical relations between cross sections,
branching ratios and coupling constants from both sectors. More
importantly we showed that the electroweak and the strong sectors
could be distinguished in a large portion of the parameter space. We
also showed that the interference between both FCNC sectors is
mostly constructive. It is also worth mentioning  that a
near-proportionality between the cross section of associated top
plus photon production at the LHC and the sum of the FCNC decays of
the top to a photon and a gluon was found. Finally, we have
estimated the backgrounds to these processes and concluded that
one might expect a significant number of events at the LHC.\\[0.2cm]

In this work we plan to extend this study to the electroweak FCNC
contributions to processes $pp \rightarrow t \, \bar{q} \, (q
\bar{q} \rightarrow t \bar{q})$, $pp \rightarrow t \, q \, (q q
\rightarrow t q)$ and complex conjugate processes and $pp
\rightarrow t \, \bar{t}$. This will finally put us in a position to
discuss FCNC single top production and decay with all the
contributions from new physics considered taken into account. Before
going any further let us stress that the choice of operators
followed a very simple rule: it had to contain one and only one top
quark and at least one gauge boson. Therefore, these operators will
always give rise to FCNC vertices contributing to the top quark FCNC
decays to a light quark and a neutral gauge boson, either strong or
electroweak. These operators have only to obey our selection
criteria to be explained in detail later - not to affect low energy
physics. How far can one go in distinguishing the different sectors
and if possible even the different operators of the theory? This is
what we are pursuing in this work - probing the actual models
proposed in the literature, and perhaps new ones, is the ultimate
goal of this entire project.

There is a guiding principle in our choice of operators in all
analysis done so far. We choose the dimension six operators that
have no sizeable impact on low energy physics. B
physics~\cite{Fox:2007in} is the main source of constraints, due
to the gauge structure of the SM. Empirically, we expect a
hierarchy of constraints resulting from the underlying SM gauge
structure. Therefore, the operators denoted by $LL$ which are the
ones built with two $SU(2)$ doublets are those where the gauge
structure is felt more strongly. On the contrary, $RR$ operators
should be the least constrained, as there is no relation between a
$R$ top quark and $R$ bottom quark. A recent study based on
constraints from B physics~\cite{Fox:2007in} has proven our
criteria to be well established. Using the LHC~\cite{toni, fla,
CMS} predictions they were able to show that, in fact, some of the
constraints on dimension 6 operators stemming from low energy
physics are already stronger than what is predicted that could be
measured at the LHC. The expected bounds obtained for a luminosity
of 100 $fb^{-1}$ are below the limits obtained for the $LL$
operators. As expected, they conclude that the $RR$ operators will
definitely be probed at the LHC, while limits on $LR$ and $RL$
operators are close to those experimental bounds. Finally, we note
that the Tevatron and the B factories are still collecting data.
Therefore the constraints will be even stronger by the time the
LHC starts to analyse data. However, it should be noted that that
are models where cancellations between different operators of type
LL could occur~\cite{delAguila:2000rc}. Those models could avoid B
physics bounds and have definitely to be analyzed in more detail
regarding the constraints from B physics.

 The use of
anomalous couplings to study possible new top physics at the LHC and
Tevatron has been the subject of many works~\cite{whis}. We just
highlight  here what are the main advantages of our approach: first
of all, it is the first time that a calculation with a complete set
of operators from both the strong and electroweak sectors is made
for single top production involving FCNC. Whenever possible and
relevant, we have presented all the expressions for the cross
sections and branching ratios allowing the use by others and in
particular in the generators available in the market. In the papers
published so far we have tried to put the emphasis on the relation
between the physical quantities, avoiding as much as possible the
use of the coupling constants related to each particular operator.
When doing so one has to keep in mind that the dependence in the
coupling constants is however still there and has to be handled with
care. Finally we note that contrary to the form factor approach, the
use of the effective operator formalism has predictive power - the
underlying gauge structure reflects itself in relations between the
several physical quantities.  We will use all available experimental
data available to restrict the parameter space. In particular we
will use the most recent data on single top production by the
CDF~\cite{CDFsingle} and D0~\cite{D0single} collaborations as well
as the data on the direct FCNC top production published by
CDF~\cite{CDFdirect}.

This paper is organised as follows: in section~\ref{sec:eff} we
review the effective operator formalism and introduce our FCNC
operators, explaining what physical criteria were behind their
choice. In section~\ref{sec:brs} we use those same Feynman rules to
compute and analyse the branching ratios of the top quark FCNC
decays. In the following two sections we study the cross sections
for production of a single top with all FCNC interactions - both
strong and electroweak - included. In section~\ref{sec:disc} we
present the integrated cross section results predicted for the
Tevatron and the LHC, and discuss the impact our FCNC operators will
have on observables such as the cross section for top + jet
production. Finally, we will draw general conclusions in
section~\ref{sec:conc}.

\section{Flavour changing effective operators}
\label{sec:eff}

The effective operator formalism of Buchm\"uller and
Wyler~\cite{buch} is based on the assumption that far below some
characteristic scale $\Lambda$ the world is well-described by the
Standard Model of particle physics. Any new physical effects not yet
observed can be accounted for with the introduction of an effective
Lagrangian with a set of new interactions to be determined
phenomenologically. Such a Lagrangian would be valid at very high
energies but, at a lower energy scale, we would only perceive its
effects through a set of effective operators of dimensions higher
than four. The form of the effective Lagrangian is independent of
the model from which it is derived and the new effective operators
are only constrained by the symmetries of low energy physics, that
is, those of the SM. This is why the effective Lagrangian approach
is ideally suited for studying possible effects of physics beyond
the SM. The effective lagrangian can be written as a series in the
energy scale, such that 
\begin{equation}
{\cal L} \;\;=\;\; {\cal L}^{SM} \;+\; \frac{1}{\Lambda}\,{\cal
L}^{(5)} \;+\; \frac{1}{\Lambda^2}\,{\cal L}^{(6)} \;+\;
O\,\left(\frac{1}{\Lambda^3}\right) \;\;\; , \label{eq:l}
\end{equation}
where ${\cal L}^{SM}$ is the SM lagrangian and ${\cal L}^{(5)}$ and
${\cal L}^{(6)}$ contain all the dimension five and six operators
which, like ${\cal L}^{SM}$, are invariant under the gauge
symmetries of the SM. The number of effective operators is obviously
infinite which means that the expansion has to be truncated at some
point. We will be working at the TeV scale (2 TeV for the Tevatron
and 14 TeV for the LHC) and therefore we choose to discard operators
of dimension higher than six, as terms of order equal or superior to
$1/\Lambda^3$ ought to be quite small, and not contribute to
TeV-scale physics. The dimension five terms in the Lagrangian will
not be considered in our analysis as they break baryon and lepton
number conservation. The list of dimension six operators is quite
vast and can be found in~\cite{buch}.

\subsection{Effective operators in the strong sector}
\noindent
In~\cite{Ferreira:2005dr,Ferreira:2006in} we have studied all
flavour changing effective operators with one top quark, some
other quark and which included a gluonic field tensor. We have
named them strong effective FCNC operators as they would change
the strong interactions. We remind that our criteria in choosing
these operators were that they contributed only to FCNC top
physics, not affecting low energy physics. In that sense,
operators that contributed to top quark phenomenology but which
also affected bottom quark physics (in the notation of
ref.~\cite{buch}, operators ${\cal O}_{qG}$) were not considered.
In accordance with our criteria there are only two dimension six
operators in the strong sector that can generate interactions with
two fermions and one gluon. Following the notation of~\cite{buch}
we write these operators as
\begin{equation}
{\cal O}_{tG\phi} =
\;\;\frac{\beta^{S}_{it}}{\Lambda^2}\,\left(\bar{q}^i_L \,
\lambda^{a} \, \sigma^{\mu\nu}\,t_R\right)\, \tilde{\phi} \, G^{a
\mu \nu} \;\; , \label{eq:op1}
\end{equation}
and
\begin{equation}
{\cal O}_{tG} = i \frac{\alpha^S_{it}}{\Lambda^2}\, \bar{u}^i_R \,
\lambda^{a} \, \gamma_{\mu}  D_{\nu} t_R \, G^{a \mu \nu} \,
 , \label{eq:op2}
\end{equation}
where the coefficients $\alpha^S_{it}$ and $\beta^{S}_{it}$ are
complex dimensionless couplings. $G^a_{\mu\nu}$ is the gluonic field
tensor, $u^i_R$ stands for a right-handed quark singlet and $q^i_L$
represents the left-handed quark. FCNC occurs because these fields
belong to the first and second generation. There are also operators,
with couplings $\alpha^S_{ti}$ and $\beta^{S}_{ti}$, where the
positions of the top and $u^î$, $q^i$ spinors are exchanged in the
expressions above. The hermitian conjugate of all operators are
obviously included in the lagrangian. These operators will originate
FCNC vertices of the form $g\,t\,\bar{u_i}$ (with $u_i
\,=\,u\,,\,c$). Due to covariant derivative acting on a quark
spinor, the operators with $\alpha^S$ couplings also contribute to
quartic vertices of the form $g\,g\,t\,\bar{u_i}$,
$g\,\gamma\,t\,\bar{u_i}$ and $g\,Z\,t\,\bar{u_i}$.
\subsection{Effective operators in the electroweak sector}
\noindent In this section we present the operators stemming from the
electroweak sector that would give rise to new FCNC interactions
involving the top quark. They would in particular contribute to the
FCNC top decays $t \rightarrow q \, Z$ and $t \rightarrow q \,
\gamma$, where $q=u,c$. First we consider the chirality flipping
operators that in a renormalizable theory are present only at
one-loop level. They are equivalent to the ones in the strong
sector, the only difference being the gluonic tensor replaced by the
U(1) and SU(2) field tensors. They can be written as
\begin{equation}
{\cal O}_{tB}= i \frac{\alpha^B_{it}}{\Lambda^2}\,\bar{u}^i_R \, \,
\gamma_{\mu} D_{\nu} t_R \, B^{\mu \nu} \;\;\; , \label{eq:op5}
\nonumber
\end{equation}
\begin{align}
{\cal O}_{tB\phi} &=
\;\;\frac{\beta^{B}_{it}}{\Lambda^2}\,\left(\bar{q}^i_L\,
\sigma^{\mu\nu}\,t_R\right)\, \tilde{\phi} \,B_{\mu\nu} \;\;\; ,
\qquad \quad {\cal
O}_{tW\phi}=\frac{\beta^{W}_{it}}{\Lambda^2}\,\left(\bar{q}^i_L\, \,
\tau_{I} \, \sigma^{\mu\nu}\,t_R\right)\, \tilde{\phi}
\,W^{I}_{\mu\nu}\;\;\; ,\label{eq:op6}
\end{align}
where $B^{\mu \nu}$ and $W^{I}_{\mu\nu}$ are the $U(1)_Y$ and
$SU(2)_L$ field tensors. The couplings $\alpha^B_{ti}$,
$\beta^{B}_{ti}$ and $\beta^{W}_{ti}$ are complex dimensionless
couplings (the couplings corresponding to the exchange of quark
spinors, and the hermitian conjugates of the operators above, were
also considered).

Expanding these operators and diagonalizing the fields' mass
matrices we obtain the terms of the form $Z \,\bar{t} \, u_i$ and
$\gamma\, \bar{t} \, u_i$. The operators considered here give no
contribution to both the gauge boson and the fermions mass matrices.
This allows us to write these operators with the fermion mass
eigenstates from the start with no loss of generality. For the same
reason, the relation between the gauge bosons mass eigenstates and
group eigenstates is the same as in the SM, i.e., they are still
related through the well-known Weinberg rotation. New effective
couplings $\{\alpha^\gamma\,,\,\beta^{\gamma}\}$ and
$\{\alpha^Z\,,\,\beta^Z\}$, related to the initial couplings via the
Weinberg angle $\theta_W$, are given by
\begin{equation}
\alpha^{\gamma}\;=\;\cos\theta_W \, \alpha^{B} \qquad \; , \;
\qquad \alpha^{Z}\;=\; - \sin\theta_W \, \alpha^{B} \label{eq:alf}
\end{equation}
and
\begin{equation}
\left\{
\begin{array}{c}
   \beta^{\gamma} \, = \, \sin\theta_W \beta^{W} + \cos \theta_W \beta^{B}\\
  \beta^{Z} \, = \, \cos\theta_W \beta^{W} - \sin \theta_W \beta^{B}  \\
\end{array}
\right. . \label{eq:bet}
\end{equation}
We showed in~\cite{Ferreira:2008cj} that the Weinberg rotation
introduces a certain correlation between FCNC processes involving
the photon or the $Z$, barring some bizarre and unnatural
cancellation between anomalous couplings in eq.~\eqref{eq:bet}.

Besides chirality-flipping operators there are chirality conserving
operators. These are tree-level operators in the sense that their
flavour conserving versions are already present in the SM at
tree-level. In fact, if we consider the vertex $\bar{t} t Z$, we see
that it has two vector contributions of different magnitudes, one
proportional to $\gamma_{\mu}\, \gamma_L$ and the other proportional
to $\gamma_{\mu}\, \gamma_R$. The anomalous operators would then
contribute to modify the Z boson neutral current. The Higgs field
has no electric charge but still interacts with the $Z$ boson. Thus,
there extra effective operators which contribute to $Z$ FCNC
interactions, analogous to those considered in~\cite{Lept} to study
FCNC in the leptonic sector. They are given by
\begin{align}
{\cal O}_{D_t} &=\frac{\eta_{it}}{\Lambda^2}\,\left(\bar{q}^i_L\,
D^{\mu}\,t_R\right)\, D_{\mu} \tilde{\phi} \, \;\;\;,\;\;\; {\cal
O}_{\bar{D}_t}=\frac{\bar{\eta}_{it}}{\Lambda^2}\,\left( D^{\mu}
\bar{q}^i_L\, \,t_R\right)\, D_{\mu} \tilde{\phi} \label{eq:ope}
\end{align}
and
\begin{equation}
{\cal O}_{\phi_t}
 \, = \, \theta_{it} \, (\phi^{\dagger} D_{\mu} \phi) \, (\bar{u^i_R} \gamma^{\mu} t_R)  \;\;\;
 ,
\label{eq:opt}
\end{equation}
and another operator with coupling $\theta_{ti}$ with the position
of the $u^i$ and $t$ spinors exchanged. As before, the coefficients
$\eta_{it}$, $\bar{\eta}_{it}$ and $\theta_{it}$ are complex
dimensionless couplings. For simplicity we redefine the $\eta$ and
$\theta$ couplings as $ \eta \rightarrow (\sin (2 \theta_W)/e) \,
\eta$ and $ \theta \rightarrow (\sin (2 \theta_W)/e) \, (\theta_{it}
\,-\,\theta^*_{ti})$. The Feynman rules for the FCNC triple vertices
are shown in the appendix. Just like for the anomalous operators in
the strong sector, the gauge structure of the terms in the
electroweak lagrangian gives rise to new quartic vertices. We will
not need those vertices in this work and refer the reader
to~\cite{Ferreira:2008cj} for details.

\section{FCNC branching ratios of the top}
\label{sec:brs}

In the SM the top decays to a light quark and a neutral gauge
boson only at the loop-level. This fact, together with the
Glashow-Iliopoulos-Maiani (GIM) mechanism, is the reason for the
immense suppression of the branching ratios of these rare top
decays in the SM. They can however be much larger in extensions of
the SM. If new particles are present, the GIM mechanism can be
avoided and, with the addition of potentially large coupling
constants from the new interactions, the branching ratios can be
enhanced by as much as thirteen orders of magnitude. For instance,
in the SM the branching ratio for the decay $t\,\rightarrow\,u\,Z$
is of the order of $\sim\,10^{-16}$, whereas in a quark-singlet
model it can reach values of as much as $10^{-4}$. For
supersymmetry or two Higgs double models, the top quark FCNC
branching ratios are typically of the order of
$10^{-6}$-$10^{-4}$. For more details
see~\cite{AguilarSaavedra:2004wm, calc}.

The effective operator formalism allows us to describe, in a
model-independent manner, the possible rare decays of the top. In
ref.~\cite{Ferreira:2005dr} we computed the branching ratios for
the FCNC top decays $t\,\rightarrow\,q\,g$, due to the strong
sector anomalous operators therein introduced. The decay width for
$t\,\rightarrow\,u\,g$ is given by
\begin{align}
\Gamma (t \rightarrow u g) &=\;  \frac{m^3_t}{12
\pi\Lambda^4}\,\Bigg\{ m^2_t \,\left|\alpha_{tu}^S  +
(\alpha^S_{ut})^* \right|^2 \,+\, 16 \,v^2\, \left(\left|
\beta_{tu}^S \right|^2 + \left| \beta_{ut}^S \right|^2 \right)
\;\;\; +
\vspace{0.3cm} \nonumber \\
 & \hspace{2.2cm}\, 8\, v\, m_t\,\mbox{Im}\left[ (\alpha_{ut}^S  + (\alpha^S_{tu})^*)
\, \beta_{tu}^S \right] \Bigg\} \label{eq:widS}\;\;\; ,
\end{align}
with an analogous expression for $\Gamma (t \rightarrow c g)$,
with different couplings. In ref.~\cite{Ferreira:2008cj} we have
repeated the calculation for the electroweak sector new FCNC
decays, namely, $t\,\rightarrow \,u\,\gamma$ (and $t\,\rightarrow
\,c\,\gamma$, with {\em a priori} different couplings), for which
we have obtained a width given by the following
expression\\[0.05cm]
\begin{align}
\Gamma (t \rightarrow u \gamma) &=\;  \frac{m^3_t}{64
\pi\Lambda^4}\,\Bigg\{ m^2_t \,\left|\alpha_{tu}^{\gamma}  +
(\alpha^{\gamma}_{ut})^* \right|^2 \,+\, 16 \,v^2\, \left(\left|
\beta_{tu}^{\gamma} \right|^2 + \left| \beta_{ut}^{\gamma}
\right|^2 \right) \;\;\; +
\vspace{0.3cm} \nonumber \\
 & \hspace{2.2cm}\, 8\, v\, m_t\,\mbox{Im}\left[ (\alpha_{ut}^{\gamma}  + (\alpha^{\gamma}_{tu})^*)
\, \beta_{tu}^{\gamma} \right] \Bigg\} \label{eq:widW} \;\;\; .
\end{align}
Notice how similar this result is to eq.~\eqref{eq:widS}. We will
also have contributions~\cite{Ferreira:2008cj} from these
operators to $t\,\rightarrow \,u\,Z$ ($t\,\rightarrow \,c\,Z$),
from which we obtain a width
given by\\[0.05cm]
\begin{eqnarray}
 \Gamma(t\,\rightarrow \,u\,Z) & = & \frac{{\left( m_t^2 - m_Z^2 \right) }^2}{32\,m_t^3\,\pi
\,\Lambda^4}
\left[ K_1 \, \left| \alpha^Z_{ut} \right|^2 + K_2 \, \left|
\alpha^Z_{tu} \right|^2 + K_3 \, ( \left| \beta^Z_{ut} \right|^2 +
\left| \beta^Z_{tu} \right|^2)+ K_4 \, ( \left| \eta_{ut}
\right|^2 + \left| \bar{\eta}_{ut} \right|^2) \right.
\nonumber \\[0.25cm]
&&  \qquad + \, K_5 \, \left| \theta \right|^2 + K_6 \, Re \left[
\alpha^Z_{ut} \, \alpha^Z_{tu} \right] + K_7 \, Im \left[
\alpha^Z_{ut} \, \beta^Z_{tu} \right]
\nonumber \\[0.25cm]
&&  \qquad + \, K_8 \, Im \left[ \alpha^{Z^*}_{tu} \, \beta^Z_{tu}
\right] + K_9 \, Re \left[ \alpha^Z_{ut} \theta^* \right]+ K_{10}
\, Re \left[ \alpha^Z_{tu} \theta \right]
\nonumber \\[0.25cm]
&& \qquad   \left. + \, K_{11} \, Re \left[ \beta^Z_{ut}
(\eta_{ut}-\bar{\eta}_{ut})^* \right] + K_{12} \, Im \left[
\beta^Z_{tu} \, \theta \right] + K_{13} \, Re \left[ \eta_{ut}
\bar{\eta}_{ut}^* \right] \right]\;\;\; ,
\end{eqnarray}
where the coefficients $K_i$ are given by\\[0.05cm]
\begin{eqnarray}
K_1 & = & \frac{1}{2} \, (m_t^4 + 4\,m_t^2\,m_Z^2 + m_Z^4) \qquad
K_2 \, = \, \frac{1}{2} \,  (m_t^2 - m_Z^2)^2 \qquad K_3 \, = \,
4\,( 2\,m_t^2 + m_Z^2) \,v^2
 \nonumber  \\
K_4 & = & \frac{v^2}{4\,m_Z^2}(m_t^2 - m_Z^2)^2 \qquad K_5 \, = \,
\frac{v^4}{m_Z^2} ( m_t^2 + 2\,m_Z^2 ) \qquad K_6 \, = \, ( m_t^2
- m_Z^2 ) \,( m_t^2 + m_Z^2)
\nonumber \\
K_7 & = & 4\,m_t\, ( m_t^2 + 2\,m_Z^2 ) \,v \qquad K_8 \, = \,
4\,m_t\, ( m_t^2 - m_Z^2 ) \,v \qquad K_9 \, = \, -2\, ( 2\,m_t^2
+ m_Z^2 ) \,v^2
\nonumber \\
K_{10} & = & -2\, ( m_t^2 - m_Z^2) \,v^2 \qquad K_{11} \, = -
K_{10} \qquad K_{12} \, = \, -12 \,m_t\,v^3 \qquad K_{13} \, = \,
\frac{- v^2 }{m_Z^2} \, K_2 \;\;\;\ .
\end{eqnarray}
\begin{table}[t]
\begin{center}
  \begin{tabular}{ | l | c | c | c |}
    \hline
     & LEP & HERA & Tevatron  \\ \hline \hline
    $Br(t \rightarrow q \, Z)$      & $ < \, 7.8 \% \,$~\cite{LEP2Zgamma} & $  < \, 49\% \,$~\cite{Zeus}
    & $ < \, 3.7 \% \,^d$~\cite{tZqCDF}  \\ \hline
    $Br(t \rightarrow q \, \gamma)$ & $ < \, 2.4 \% \,$~\cite{LEP2Zgamma} & $ < \, 0.75 \% \,$~\cite{Zeus}
    & $ < \, 3.2 \% \,^d$~\cite{gammaCDF}  \\ \hline
    $Br(t \rightarrow q \, g)$      & $ < \, 17 \% \,$~\cite{YR1}  & $ < \, 13 \% \,$~\cite{Ashimova:2006zc,Zeus}
    & $ < \, O (0.1 - 1 \%) \,$~\cite{gluonTevatron,Lee} \\
    \hline
  \end{tabular}
\end{center}
\caption{Current experimental bounds on FCNC branching ratios. The
superscript ``d" refers to bounds obtained from direct
measurements, as is explained in the text.} \label{tab:limits}
\end{table}
\noindent FCNC processes were the subject of searches by several
experimental groups. Indirect bounds~\cite{Fox:2007in,EPM} are
provided mostly by electroweak precision physics and B and K
physics. Due to the SM gauge structure, B physics can be used to set
limits on operators that involve top and bottom quarks. The
strongest bounds so far are the ones in \cite{Fox:2007in} where
invariance under $SU(2)_L$ is required for the set of operators
chosen. Regarding $Br (t \,\rightarrow\, q\, Z)$ and $Br
(t\,\rightarrow \,q\, \gamma)$, the only direct bounds available to
date are the ones from the Tevatron (CDF). The CDF collaboration has
searched its data for signatures of $t \,\rightarrow \,q \, \gamma$
and $t\, \rightarrow \,q \, Z$ (where $q\,=\,u,c$). Both analyses
use $p\bar{p}\, \rightarrow \, t\,\bar{t} $ data and assume that one
of the tops decays according to the SM into $W\,b$. We have
collected all these experimental results in Table \ref{tab:limits}.
LEP and ZEUS have translated to bounds on the branching ratios,
experimental limits originally obtained for the cross section. The
translation is straightforward as it was done via just one operator,
the chromomagnetic one. Note that LEP bounds are derived assuming
the same anomalous coupling for the $u$ and $c$ quarks while the
ZEUS bound is only for the process involving a $u$ quark. With
forthcoming Tevatron data, we expect that these bounds will improve
in the near future. The same searches are being prepared for the
LHC. A detailed discussion with all present bounds on FCNC and the
predictions for the LHC can be found in \cite{toni, fla, CMS}. With
a luminosity of 100 $fb^{-1}$ and in the absence of signal, the 95\%
confidence level bounds on the branching ratios give us $Br(t
\,\rightarrow\, q\, Z)\,\sim\,10^{-5}$, $Br(t \,\rightarrow\, q\,
\gamma)\,\sim\,10^{-5}$ and $Br(t \,\rightarrow\, q\,
g)\,\sim\,10^{-4}$.

\section{Electroweak FCNC contributions to top + quark production}

In previous references~\cite{Ferreira:2006xe},  we calculated the
FCNC cross sections for several processes of single top production.
To wit, all of the strong-FCNC contributions to the processes $pp
\rightarrow gq \rightarrow gt$, $pp \rightarrow gg \rightarrow
\bar{q}t$, $pp \rightarrow qq \rightarrow qt$, and respective
charged conjugate processes. This amounted to a complete calculation
for $pp \rightarrow t+jet$ in the strong sector. Now, in
reference~\cite{Ferreira:2008cj} it was shown that the electroweak
FCNC interactions can have contributions to processes of associated
top production alongside a neutral gauge boson as large as the
strong ones. Thus, in order for us to have the complete FCNC
contribution to the cross section for top + jet production, it
becomes imperative to compute the electroweak FCNC contributions to
this process, and their interference with the strong FCNC vertices.
It is of great interest to possess a complete FCNC expression for
this cross section, as it is one of the most basic top quark
observables that can be measured in colliders.

Obviously, the electroweak sector will give no contribution to
processes with gluons in the initial or final states. Only the
processes presented in table~II will be modified, which constitute
all possibilities for the single top channel
$q\,q\,\rightarrow\,t\,q$.
\begin{table}[h!]
\begin{center}
\begin{tabular}{cc}\hline\hline \\
 Single top channel & Process number \\ & \\ \hline \\
$u\,u\,\rightarrow\,t\,u$  & 1 \vspace{0.2cm} \\
$u\,c\,\rightarrow\,t\,c$  & 2 \vspace{0.2cm}\\
$u\,\bar{u}\,\rightarrow\,t\,\bar{u}$  & 3 \vspace{0.2cm}\\
$u\,\bar{u}\,\rightarrow\,t\,\bar{c}$  & 4 \vspace{0.2cm}\\
$u\,\bar{c}\,\rightarrow\,t\,\bar{c}$  & 5 \vspace{0.2cm}\\
$d\,\bar{d}\,\rightarrow\,t\,\bar{u}$  & 6 \vspace{0.2cm}\\
$u\,d\,\rightarrow\,t\,d$  & 7 \vspace{0.2cm}\\
$u\,\bar{d}\,\rightarrow\,t\,\bar{d}$  & 8 \\ &
\\\hline\hline\hline
\end{tabular}
\caption{List of single top production channels through
quark-quark scattering.}
\end{center}
\end{table}
The processes in table II should be divided in two groups. Processes
6 to 8 are present in the SM at tree-level while processes 1 to 5
appear only at the loop-level. The first new terms for processes 6
to 8 are therefore the interference between the SM term and the
order $1/\Lambda^2$ contributions from the effective Lagrangian. For
channels 1 to 5 the first term is the square of the term of order
$1/\Lambda^2$ in the Lagrangian. Naively, one would expect terms 6
to 8 to be much larger than the 1 to 5 ones because of their
dependence in the scale of new physics. However, due to a very
strong cancellation caused by the Cabbibbo-Kobayashi-Maskawa (CKM)
matrix elements, it turns out that the interference between the
tree-level SM processes and the anomalous ones are extremely small.
Hence, contrary to what we would expect, the anomalous contributions
at the lowest order in $\Lambda$ are not the best ones to look for
FCNC physics beyond the SM, at least in this channel. Processes 1 to
5, although of order $1/\Lambda^4$, are indeed much larger than the
last ones, since in this case no CKM cancellation takes place.
Finally, we remind that there are $1/\Lambda^4$ contributions to
processes 6 to 8 as well. However, terms of dimension eight in the
effective Lagrangian would interfere with the SM tree-level process
to give terms of that same order. According to the philosophy we
have adopted, we decided not to take them into account as the
calculation would not be complete.

\begin{figure}[h!]
  \begin{center}
    \epsfig{file=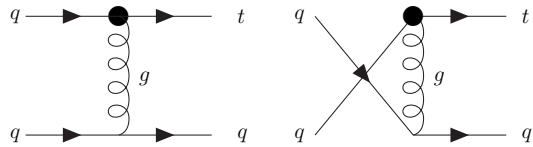,width=8 cm}
    \caption{Feynman diagrams for strong FCNC $t \, q$ production.}
    \label{fig:fig_qq_qt}
  \end{center}
\end{figure}

\begin{figure}[h!]
  \begin{center}
    \epsfig{file=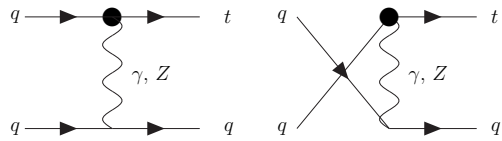,width=8 cm}
    \caption{Feynman diagrams for electroweak FCNC $t \, q$ production.}
    \label{fig:fig_qq_qtweak}
  \end{center}
\end{figure}

\begin{figure}[h!]
  \begin{center}
    \epsfig{file=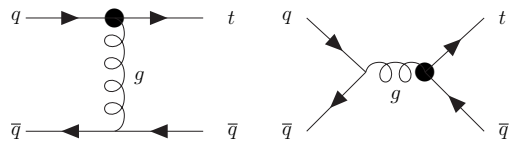,width=8 cm}
    \caption{Feynman diagrams for strong FCNC $t \, \bar{q}$ production.}
    \label{fig:fig_qqbar_qtbar}
  \end{center}
\end{figure}

\begin{figure}[h!]
  \begin{center}
    \epsfig{file=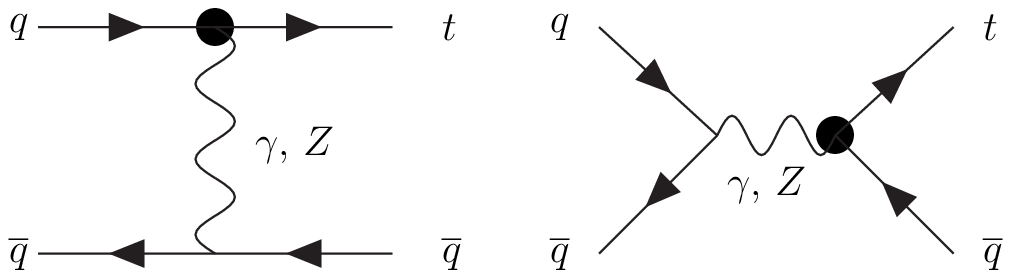,width=8 cm}
    \caption{Feynman diagrams for electroweak FCNC $t \, \bar{q}$ production.}
    \label{fig:fig_qqbar_qtbarweak}
  \end{center}
\end{figure}

In figures~\eqref{fig:fig_qq_qt} and \eqref{fig:fig_qq_qtweak} we
show the strong and electroweak contributions, respectively, to the
process $pp \rightarrow qq \rightarrow qt$.
Both processes proceed through a t and a u channel. In
figures~\eqref{fig:fig_qqbar_qtbar} and
\eqref{fig:fig_qqbar_qtbarweak} we show the strong and electroweak
contributions, respectively, to the process $pp \rightarrow q
\bar{q} \rightarrow t \bar{q}$.
Note that we only consider processes where there is a ``single"
flavour violation, that is a single FCNC vertex per diagram. Now, in
the work of
refs.~\cite{Ferreira:2005dr,Ferreira:2006xe,Ferreira:2006in,Ferreira:2008cj}
we endeavoured to present analytical expressions for all cross
sections and decay widths computed, to make them available to
whomever might wish to use them. For the full $t\,+\,q$ cross
section expression, however, that is no longer practical, as the
formulae are extremely lengthy. For the reader to appreciate the
level of complexity of the results we obtained, suffice it to say
that the full cross section is a sum of 66 different combinations of
anomalous couplings, namely
\begin{align}
\frac{d\sigma}{dt} & =
F_1\,|\alpha^S_{ut}|^2+\,F_2\,|\alpha^S_{tu}|^2\,
+\,F_3\,\Bigl[|\beta^S_{ut}|^2\,+\,|\beta^S_{tu}|^2\Bigr] \, +\,F_4
\, Re(\alpha^S_{ut}\, \alpha^S_{tu})\,+\,F_5 \, Im(\alpha^S_{ut}\,
\beta^{S}_{tu}) \vspace{0.2cm} \nonumber\\
& \;\;+\,F_6 \, Im(\alpha^S_{tu}\, \beta^{S*}_{tu}) \,+\,
G_1\,|\alpha^\gamma_{ut}|^2+\,G_2\,|\alpha^\gamma_{tu}|^2\,
 +\,G_3\,\Bigl[|\beta^\gamma_{ut}|^2\,+\,|\beta^\gamma_{tu}|^2\Bigr] \,
  +\,G_4 \, Re(\alpha^\gamma_{ut}\, \alpha^\gamma_{tu})\vspace{0.2cm} \nonumber\\
 & \;\; +\,G_5 \,
Im(\alpha^\gamma_{ut}\, \beta^{\gamma}_{tu})\,+\,G_6 \,
Im(\alpha^\gamma_{tu}\, \beta^{\gamma *}_{tu}) \,+\,
H_1\,|\alpha^Z_{ut}|^2+\,H_2\,|\alpha^Z_{tu}|^2\,
+\,H_3\,|\beta^Z_{ut}|^2 \vspace{0.2cm} \nonumber\\
 & \;\; +\,H_4 \,
|\beta^Z_{tu}|^2\,+\,H_5
\,\Bigl[|\eta|^2\,+\,|\bar\eta|^2\,-2\,Re(\eta\, \bar\eta^*)\Bigr]
\,+\,H_6 \, |\theta|^2 \,+\,H_7 \, Re(\alpha^Z_{ut}\, \alpha^Z_{tu})
\,+ H_8 \, Im(\alpha^Z_{ut}\, \beta^Z_{tu}) \vspace{0.2cm} \nonumber\\
 & \;\; + \, H_9 \,
Re(\alpha^Z_{ut}\, \theta^*) \, + \, H_{10} \, Im(\alpha^Z_{tu}\,
\beta^{Z*}_{tu})  \, + \, H_{11} \, Re(\alpha^Z_{tu}\,
\theta)\,+\,H_{12} \Bigl[ Re(\beta^Z_{ut}\, \eta^*) \,-
\,Re(\beta^Z_{ut}\, \bar \eta^*) \Bigr] \vspace{0.2cm} \nonumber\\
 & \;\;+ \, H_{13} \,
Im(\beta^Z_{tu}\, \theta) \,+\, FG_{1}\,Re(\alpha^{S}_{ut}\,
\alpha^{\gamma *}_{ut})\,+\, FG_{2}\,\Bigl[ Re(\alpha^{S}_{ut}\,
\alpha^{\gamma }_{tu})\, + \, Re(\alpha^{S}_{tu}\, \alpha^{\gamma
}_{ut})  \, - \, Re(\alpha^{S}_{tu}\, \alpha^{\gamma *}_{tu}) \Bigr]\vspace{0.2cm} \nonumber\\
 & \;\; +\,FG_{3}\,\Bigr[Im(\alpha^{S}_{ut}\, \beta^{\gamma}_{tu}) \, + \,
Im(\beta^{S}_{tu}\, \alpha^{\gamma}_{ut}) \Bigr]\,+\,FG_{4}\,\Bigl[
Re(\beta^{S}_{ut}\, \beta^{\gamma *}_{ut})\, + \,
Re(\beta^{S}_{tu}\, \beta^{\gamma *}_{tu})  \Bigr]\vspace{0.2cm} \nonumber\\
 & \;\; +\,FH_1\,Re(\alpha^S_{ut}\, \alpha^{Z *}_{ut}) \,+ \,
 FH_2\,\Bigl[Re(\alpha^S_{ut}\, \alpha^{Z}_{tu})\, + \, Re(\alpha^S_{tu}\, \alpha^{Z}_{ut}) \,
 - \,Re(\alpha^S_{tu}\, \alpha^{Z *}_{tu})\Bigr] \vspace{0.2cm} \nonumber\\
 & \;\; +\,FH_3\Bigl[ Im(\alpha^S_{ut}\, \beta^{Z}_{tu}) \, + \, Im(\beta^S_{tu}\, \alpha^{Z}_{ut})\Bigr]
\, +\,FH_4 \, Re(\alpha^S_{ut}\, \theta^*)\,+\,FH_5 \,Re(\alpha^S_{tu}\, \theta) \vspace{0.2cm} \nonumber\\
 & \;\; + \,FH_6 \, Re(\beta^S_{ut}\, \beta^{Z *}_{ut}) +\,FH_7 \Bigl[Re(\beta^S_{ut}\, \eta^*) \, - \,
 Re(\beta^S_{ut}\, \bar\eta^*)\Bigr] \,+ \,FH_8 \, Re(\beta^S_{tu}\, \beta^{Z *}_{tu}) \vspace{0.2cm} \nonumber\\
 & \;\;  + \, FH_9 \,
Im(\beta^S_{tu}\, \theta) \,+\,GH_1\,Re(\alpha^\gamma_{ut} \, \alpha^Z_{ut})\, + \,GH_2\Bigl[Re(\alpha^\gamma_{ut} \, \alpha^Z_{tu}) \, + \, Re(\alpha^\gamma_{tu} \, \alpha^Z_{ut})\Bigr] \vspace{0.2cm} \nonumber\\
 & \;\;  +\,GH_3\Bigl[Im(\alpha^\gamma_{ut} \, \beta^Z_{tu}) \, + \, Im(\beta^\gamma_{tu} \, \alpha^Z_{ut})\Bigr]
 \, +\,GH_4 \, Re(\alpha^\gamma_{ut} \, \theta)\vspace{0.2cm} \,  +\,GH_5 \,Re(\alpha^\gamma_{tu} \, \alpha^{Z\,*}_{tu}) \nonumber\\
 & \;\;+\,GH_6\Bigl[Im(\alpha^\gamma_{tu} \, \beta^{Z\,*}_{tu}) \, + \, Im(\beta^\gamma_{tu} \, \alpha^{Z\,*}_{tu})\Bigr]\,+\,GH_7 \, Re(\alpha^\gamma_{tu} \, \theta)  + GH_8 \, Re(\beta^\gamma_{ut} \, \beta^{Z\,*}_{ut}) \nonumber\\
 & \;\;+ \, GH_9 \Bigl[Re(\beta^\gamma_{ut} \, \eta^{Z\,*}) \, - \, Re(\beta^\gamma_{ut} \, \bar\eta^{Z\,*})\Bigr] \,+ \, GH_{10} \, Re(\beta^\gamma_{tu} \, \beta^{Z\,*}_{tu})  \, + \,
GH_{11} \, Im(\beta^\gamma_{tu} \, \theta) \;\;\; ,
 \end{align}
 where $F_i$, $G_i$, \ldots are very lengthy functions of the Mandelstam
 variables $s$ and $t$, and of the masses of the particles present.
 For the interested reader, the authors the full
 expressions in Mathematica format are available in
 http://mars.fis.uc.pt/$\sim$plhc-top/public/publications.shtml. See
 also~\cite{miguel}. These expressions were obtained with the
help of the FeynCalc package~\cite{Mertig:1990an}.

\begin{figure}[h!]
  \begin{center}
    \epsfig{file=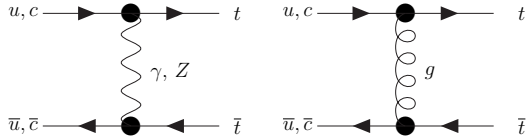,width=8 cm}
    \caption{Feynman diagrams for FCNC $t \, \bar{t}$ production.}
    \label{fig:fig_qq_ttbar}
  \end{center}
\end{figure}

Another interesting calculation that this full set of FCNC operators
allows us to do is the contribution to the production of $t\bar{t}$
pairs at the LHC. This arises from the interference of the SM
tree-level diagrams and diagrams with {\em two} anomalous vertices.
Thus, the resulting cross sections are, once again, of the order
$1/\Lambda^4$. In fact, in~\cite{Ferreira:2006in} we had calculated
the interference between the SM processes
$q\,\bar{q}\,\rightarrow\,t\,\bar{t}$ and
$g\,g\,\rightarrow\,t\,\bar{t}$ and the double FCNC diagrams from
the strong sector. We now completed the calculation, with the
contributions from both the electroweak and the strong sectors as
shown in the diagrams of figure~\eqref{fig:fig_qq_ttbar}.
As with the cross section for top + jet production, the final
expressions are extremely cumbersome and will not be presented here,
but are available upon demand. Finally, we did not study the
production of top quarks of the same sign, because once again terms
of  dimension eight in the effective Lagrangian would have to be
taken into account for a consistent and complete description.

\section{Top production via FCNC interactions at the Tevatron and LHC}
\label{sec:disc}

With the full cross sections computed, the anomalous operators
considered here contribute to the following physical processes: $pp
\rightarrow t j$, $pp \rightarrow t \bar{t}$, $pp \rightarrow t Z$,
$pp \rightarrow t \gamma$ and $pp \rightarrow t h$, where $j$ is a
jet and $h$ is the Higgs boson. All these processes were the subject
of detailed studies in previous papers. However, this is the first
time that both strong and electroweak operators will be used in the
analysis of the process $pp \rightarrow t j$. As defined by our
criteria in the introduction, the study of the effective FCNC
operators with one top quark and at least one gauge boson is now
complete.

The cross sections derived here may be applied to any collider we
wish to study. In particular, as we will shortly see, the anomalous
FCNC interactions here considered would already have a sizeable
impact on Tevatron top physics, if the top FCNC branching ratios
were large. We are mostly interested in LHC physics, so our
philosophy will be to use all available data from the Tevatron
experiments to constrain the values of the anomalous couplings. This
will enable us to curtail the available parameter space and refine
our LHC predictions. As we will see, the Tevatron data has a
substantial impact on what one might expect for FCNC single top
production at the LHC.

We remind the reader that, for processes of single top production
via FCNC, since there are no interferences with the
SM~\footnote{Or those interferences are exceedingly small, as
mentioned regarding the processes 6, 7 and 8 from table II.}, our
FCNC cross sections constitute {\em extra} contributions to the SM
cross sections. In other words, our anomalous cross sections add
to the expected SM values. In table~\ref{tab-singl} we present the
SM predictions for the single top cross sections at the LHC,
against which the size of our FCNC contributions must be compared.
\begin{table}[h!]
\begin{center}
\renewcommand{\arraystretch}{1.2}
\begin{tabular}{llll}
\hline \hline
Cross Section  & $t$-channel & $s$-channel & $tW$ mode \\
\hline $\sigma_{\rm LHC}^t$  & $150 \pm 6$  pb & $7.8 \pm 0.7$ pb
&
$44 \pm 5$  pb \\
$\sigma_{\rm LHC}^{\bar t}$  &
$92\pm 4$  pb & $4.3 \pm 0.3$ pb & $44\pm 5$ pb  \\
\hline \hline
\end{tabular}
\end{center}
\caption{\label{tab-singl} Single top-quark production cross
  sections predictions for the LHC.  The errors
include scale uncertainties, parton density function
uncertainties, and uncertainties in the top mass. The value
$m_t=171.4 \pm 2.1$ GeV was used. See
refs.~\cite{Bernreuther:2008ju} for details.}
\end{table}

A brief summary of our procedures: we generated a large set of
complex random values for the anomalous couplings. The range of
values chosen for each of the coupling constants was $10^{-12} <
|a/\Lambda^2| < 1$, where $a$ stands for a generic coupling and
the scale of new physics, $\Lambda$, is in TeV. For an excellent
discussion on the order of magnitude of the anomalous couplings
see~\cite{Wudka:1994ny}. We discard those combinations of values
of the couplings for which the several FCNC branching ratios we
computed earlier are larger than $10^{-2}$. This is a conservative
approach, since $10^{-2}$ is below all experimental upper bounds
shown in table I. Only the Tevatron bound on the strong FCNC decay
can be smaller, but this is incorporated in the analysis. When an
acceptable combination of values is found, we then use it to
compute the cross sections for the various processes. All cross
sections were obtained by integrating the corresponding partonic
cross section with the CTEQ6M partonic density
functions~\cite{cteq6}, with a factorization scale $\mu_F$ set
equal to $m_t$. We also imposed a cut of 15 GeV on the $p_T$ of
the final state partons.
\begin{figure}[h!]
  \begin{center}
    \epsfig{file=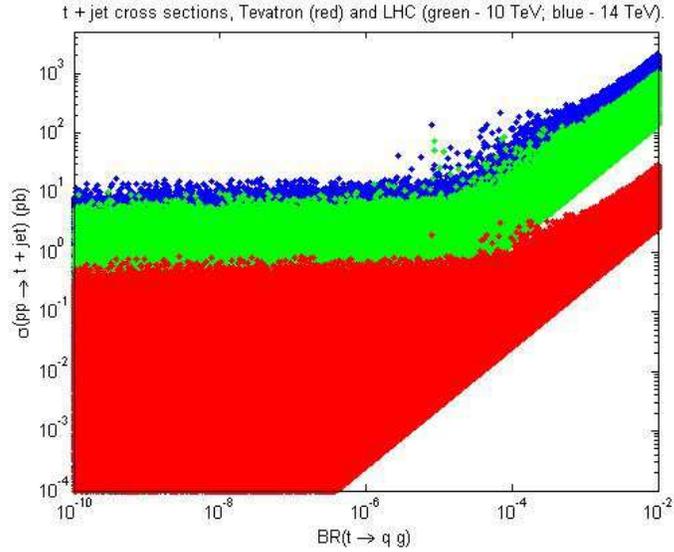,width=10 cm}
    \caption{Total FCNC contributions to the cross section of top + jet
    production as a
  function of the sum of the gluon FCNC branching ratios, for the
    Tevatron (red)and LHC (green - 10 TeV; blue - 14 TeV). }
    \label{fig:tj10}
  \end{center}
\end{figure}
\begin{figure}[h!]
  \begin{center}
    \epsfig{file=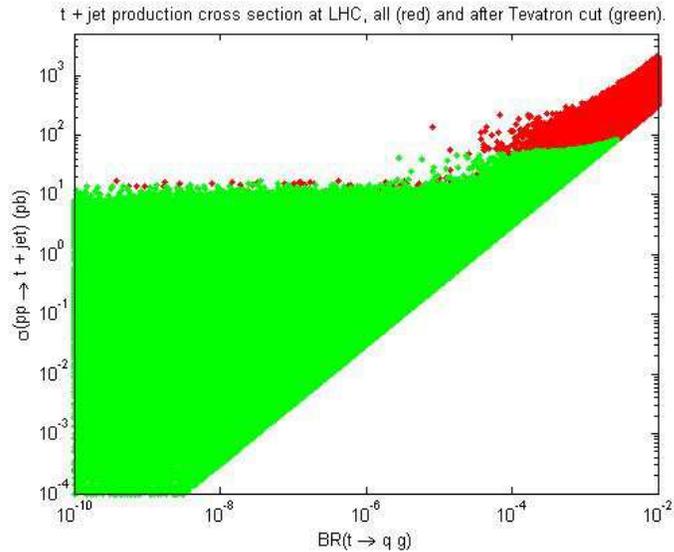,width=10 cm}
    \caption{Total FCNC contributions to the cross section of top + jet production at the LHC as a
    function of the sum of all strong FCNC branching ratios. The points which survive the Tevatron cut
    are shown in green.}
    \label{fig:alltjvsbrweak}
  \end{center}
\end{figure}

In figure~\eqref{fig:tj10} we plot the FCNC anomalous contribution
to the value of the total cross section for $t + jet$ against the
sum of branching ratios of the FCNC decay of the top to a gluon and
a $u$ quark plus the decay to a gluon and a $c$ quark. We show the
cross section values for the Tevatron, for the LHC at a low
operating energy of 10 TeV and for its nominal center-of-mass energy
of 14 TeV. This result includes direct top production and the
production of a top quark alongside a light quark or a gluon,
computed
in~\cite{Ferreira:2005dr,Ferreira:2006xe,Ferreira:2006in,Ferreira:2008cj},
as well as the new electroweak sector contributions, that is
\begin{eqnarray}
\sigma (pp\rightarrow tj) & = & \sigma^S (pp \rightarrow gq
\rightarrow t) \,+\, \sigma^S (pp \rightarrow gg \rightarrow t
\bar{q}) \,+\, \sigma^S (pp \rightarrow g q \rightarrow g t) \,+\,
\vspace{0.2cm}\nonumber \\
&&\sigma^S (pp \rightarrow q \bar{q} \rightarrow t \bar{q}) \,+\,
\sigma^{EW} (pp \rightarrow q \bar{q} \rightarrow t \bar{q}) \,+\,
\sigma^{INT} (pp \rightarrow q \bar{q} \rightarrow t \bar{q})\,+\,
\vspace{0.2cm}\nonumber \\
&&\sigma^S (pp \rightarrow q q \rightarrow t q) \,+\, \sigma^{EW}
(pp \rightarrow q q \rightarrow t q) \,+\, \sigma^{INT} (pp
\rightarrow q q \rightarrow t q) \;\;\;\ ,
\end{eqnarray}
where the superscript $S$ stands for strong contribution, $EW$ for
electroweak and $INT$ for the interference terms between the strong
and the electroweak contributions. As is plain to see from
fig.~\eqref{fig:tj10}, the FCNC contributions to single top
production can already be quite large at the Tevatron, so the
results from that collider need to be taken into account. We
emphasize that the same anomalous couplings are used to compute the
cross sections at any center of mass energy. There is also little
difference between the LHC cross sections at 10 or 14 TeV, though of
course the higher energy corresponds to larger values. From now on
all LHC results will correspond to $\sqrt{s}\,=\,14$ TeV.

Let us look more carefully at the LHC predictions, by imposing the
existing experimental constraints from the observed cross sections
of single top production at the Tevatron: the calculated FCNC
contribution to the cross section for single top at the Tevatron was
forced to be smaller than the experimental error obtained in the
latest single top measurement of the $s+t$ channels,
$\sigma_{s+t}^{Tevatron} = 2.2 \pm 0.7$ pbarn~\cite{CDFsingle} (see
also \cite{D0single}). By looking at
figure~\eqref{fig:alltjvsbrweak}, it is clear that the Tevatron
constraints make a significant impact on what can be probed at the
LHC. With one million points generated, the highest value for the
strong branching ratio is found to be around $0.3 \%$. Also, the
allowed values for the cross sections drop from a few thousand pbarn
to less than 100 pbarn. Nevertheless, the LHC will still be able to
further probe this process in the first years of running.

Another very important point worth mentioning is that, even when the
strong branching ratio becomes negligible, we still have plenty of
points above the 10 pbarn line. In other words, a modification in
the electroweak sector related to FCNC top interactions can be
sensed in an apparently strong process such as $pp \rightarrow tj$.
Note that the largest anomalous FCNC contributions to top plus jet
production at the LHC come from processes with gluons in the initial
state, and therefore would correspond to an alteration of the strong
interaction. A measurement of an excess of a few pbarn in this cross
section does not necessarily means new physics coming from the
strong sector - it could mean new physics with its origin in the
electroweak sector. An experimental limit in this cross section is
automatically translated in a limit for the strong branching ratio.
The reverse, however, is not true - an experimental limit in the
strong branching ratio will tell us nothing about the cross section
of top + jet production, due to the inclusion of the contributions
from the electroweak sector.

\begin{figure}[h!]
  \begin{center}
    \epsfig{file=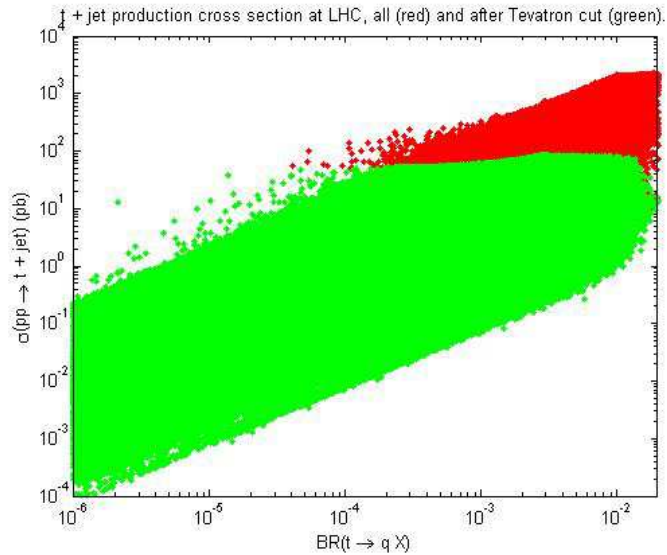,width=10 cm}
    \caption{Total FCNC contributions to the cross section of top + jet
    production at the LHC as a function of the sum of all
    FCNC branching ratios (strong and electroweak). The points which survive the Tevatron cut
    are shown in green.}
    \label{fig:alltjvsbrstrong}
  \end{center}
\end{figure}
In figure~\eqref{fig:alltjvsbrstrong} we again show the extra FCNC
contribution to the cross section for top plus jet production in
pbarn, but now against the sum of all FCNC branching ratios, that
is, $Br(t \rightarrow q g)+Br(t \rightarrow q \gamma)+ Br(t
\rightarrow q Z)$, where a sum in $q=u,c$ is implicit. With over one
million points generated, we found $Br (t \rightarrow q Z) < 1.4 \%$
and $Br (t \rightarrow q \gamma) < 1.8 \%$ for the electroweak
branching ratios, after the Tevatron constraints were imposed. These
constraints on the branching ratios are 1-sigma values (68\% C.L.).
Once again, the total cross section is shown in red, and in green we
show the values for the cross section after the constraints from the
Tevatron were imposed. Once more we see that the cross section is
bound to be below 100 pbarn, and that an experimental bound on the
branching ratio can be translated into a bound on the cross section.
It is interesting to note that even for a sum of the branching
ratios below $10^{-5}$, which is well below the sensitivity expected
for the LHC, we can still have cross sections of a few pbarn.
Therefore, an excess in the total single top cross section has to be
interpreted with great care.

\begin{figure}[h!]
  \begin{center}
    \epsfig{file=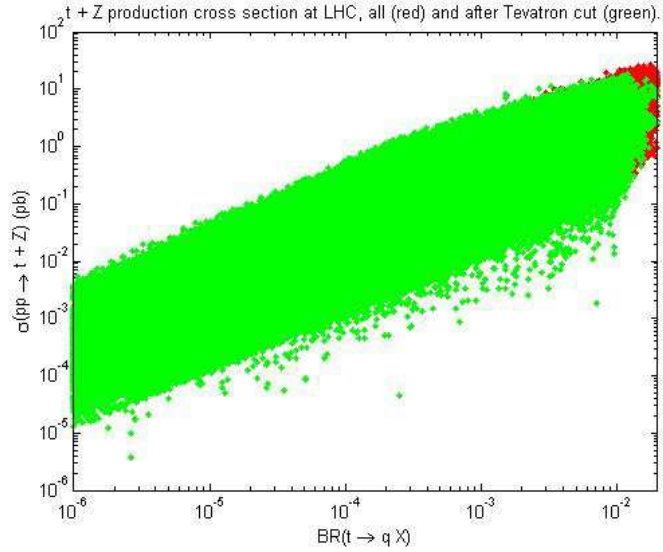,width=10 cm}
    \caption{FCNC cross section for top + Z production at the LHC as a function of
    the sum of all FCNC branching ratios (strong+electroweak). The points which survive
    the Tevatron cut
    are shown in green.}
    \label{fig:alltjvsallbrsum}
  \end{center}
\end{figure}
In figure~\eqref{fig:alltjvsallbrsum} we show the FCNC cross section
for top plus Z production at the LHC, against the total FCNC
branching ratios. As we can see, for this observable the Tevatron
cut produces no sizeable exclusion region, and the maximum value of
the cross section amounts to about $\sim 10$ pb. Nevertheless, there
is a small region for the larger values of the branching ratio that
the Tevatron has managed to exclude. The corresponding plot for the
FCNC cross section for $t + \gamma$ production is not shown here,
since the constraints from the Tevatron are even milder and have no
bearing on the final result. Moreover, the total $t + \gamma$ cross
section is found to be even smaller than that for $t + Z$.

\begin{figure}[h!]
  \begin{center}
    \epsfig{file=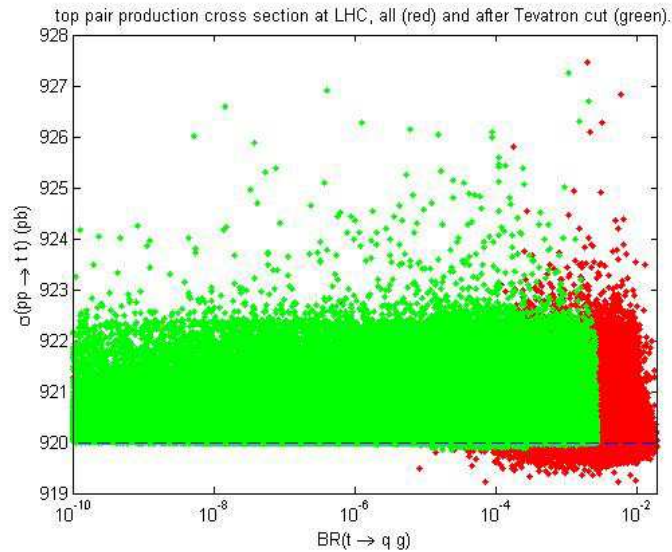,width=10 cm}
    \caption{Total $t \bar{t}$ production cross section at the LHC
    as a function of the sum of the strong FCNC branching ratios.
    The dashed line corresponds to the expected SM vale, the red points are
    excluded using Tevatron single-top data.}
    \label{fig:tt}
  \end{center}
\end{figure}
Finally, in figure~\eqref{fig:tt} we plot the total cross section
for $t \bar{t}$ production at the LHC - by ``total" we mean
including both the SM result and the anomalous FCNC contributions to
it. These latter contributions result, as was explained earlier,
from interferences between SM diagrams and FCNC ones. It is crucial
to verify the impact that our single-top FCNC operators may have on
double top production, as a consistency check - if we are changing
top quark physics in one given chanel, what are the consequences in
others? Thus, the $t \bar{t}$ cross section results already existing
could, in principle, be used to constrain single top FCNC operators.
However, we have checked that whilst varying the anomalous coupling
constants in the range described above, the FCNC extra contribution
to the $t \bar{t}$ cross section cannot reach more than 0.64 pbarn
at the Tevatron, and 7.3 pbarn at the LHC.

We observe this feature in figure~\eqref{fig:alltjvsallbrsum}, where
we used a value of $920$ pbarn to the expected SM $t \bar{t}$ cross
section at LHC~\cite{Cacciari:2008zb}. In red we show the total
cross section, while in green we present the cross section after the
Tevatron cuts. The central line is the expected SM cross section at
the LHC at 14 TeV. It is interesting to note that the Tevatron data
seems to eliminate the destructive interference with the FCNC
contributions, and the remaining anomalous contributions in the LHC
always increase the $t \bar{t}$ cross section, although not by much.
It is plain to see from this plot that even the largest values
attained for the LHC cross section for $p\,p\, \rightarrow
\,t\,\bar{t}$ will be well below the expected experimental error for
this process. Hence, due to the very small values obtained, the $t
\bar{t}$ production cross section gives us no information on the set
of FCNC single top operators. Only a future and more precise
measurement of this cross section will enable us to use the $t
\bar{t}$ data to constrain FCNC single top physics. Note, however,
that double and single top processes are already related. The
constraints on the FCNC branching ratios are obtained from $t
\bar{t}$ production with one top decaying to $bW$ and the other to
$qZ$, $q\gamma$ or $qg$.

\section{Conclusions}
\label{sec:conc}

We have performed a general analysis of the impact of dimension
six FCNC effective operators on top quark physics. The electroweak
contributions presented in this work complete the impact the set
of operators chosen might have on channels of single and double
top production. The criteria behind that selection are extremely
general: that they do not ``spoil" the agreement with physics
below the TeV scale, and that FCNC is observed in the interactions
of the top with gauge bosons. The inclusion of these new
contributions allows us to finally compute the full FCNC
contributions to an extremely important physical observable: the
cross section for single top + jet production. Our results are
trivially applied to both the Tevatron and the LHC, and we have
shown that our cross sections already predict substantial
contributions to Tevatron cross sections. In fact, we used that
fact to constrain our allowed space of anomalous constants: by
requiring that our predictions for the Tevatron be below the
experimental error for top + jet production, we excluded a
sizeable portion of parameter space available to be probed at the
LHC. Further, this allowed us to set upper bounds on several FCNC
branching ratios, which we summarise in table IV, both at the 68\%
and 95\% confidence level. Notice that there is no difference
between the bounds obtained for the electroweak
\begin{table}[h!]
\begin{center}
\renewcommand{\arraystretch}{1.2}
\begin{tabular}{ccc}
\hline \hline
Branching ratio  & Upper bound & Upper bound\\
 & (68\% C.L.) & (95\% C.L.) \\
\hline $BR(t\,\rightarrow\,q\,g)$  & 0.3\,\% & 0.55\,\% \vspace{0.2cm}\\
$BR(t\,\rightarrow\,q\,Z)$  & 1.34\,\%  & 1.34\,\% \vspace{0.2cm}\\
$BR(t\,\rightarrow\,q\,\gamma)$  & 1.77\,\%  & 1.77\,\% \vspace{0.2cm}\\
$BR(t\,\rightarrow\,q\,X)$  & 2.15\,\%  & 2.77\,\% \\
\hline \hline
\end{tabular}
\end{center}
\caption{Upper bounds for top quark FCNC branching ratios obtained
after imposing the constraints stemming from current Tevatron
results. $BR(t\,\rightarrow\,q\,X)$ is the sum of all possible FCNC
branching ratios.} \label{tab:brs}
\end{table}
FCNC branching ratios. This is due to the fact that, as had been
already observed regarding figure~\ref{fig:alltjvsallbrsum}, the
Tevatron bounds have little effect on electroweak processes.

Finally, we proved that the FCNC interactions considered had little
impact on $t\bar{t}$ produtcion. In fact, this conclusion highlights
the importance of the single top channel - top FCNC interactions, if
indeed exist, as is predicted by many an extension of the SM, will
manifest themselves in  single top physics, not top pair production.
The single top channel is thus an exquisite place to look for
signals of new physics.

The electroweak contributions were first studied in
ref.~\cite{Ferreira:2008cj} concerning their impact on channels of
top + Z or top + photon production. The emphasis on that work was on
the possibility of using physical observables to distinguish between
strong and electroweak FCNC contributions. In what regards top + jet
cross sections, we may ask whether it is also possible to use these
channels to distinguish between the strong and electroweak FCNC
contributions. However, notice that the diagrams for top + jet
production from the strong sector and from the electroweak sector
have exactly the same structure - whereas for top + Z or top +
photon production, the diagrams stemming from the strong FCNC
interactions had a strong $t$-channel contribution, and the
electroweak ones a strong $s$-channel component. This difference in
structure of both types of diagrams made it possible, through the
analysis of the differential cross sections, to, at least in
principle, individuate strong and electroweak cross sections. For
the top + jet cross sections, however, the differential cross
sections for both sectors are similar. As such, we do not expect
that this observable could be used to separate electroweak FCNC
physics from that of the strong sector.

The cross sections calculated in this paper demonstrate in a clear
manner the importance of the single top channels for studies of FCNC
physics. In some cases, the best expected experimental limit will
probably come from the study of this channel, in particular at the
LHC~\cite{Lee}. It should be stressed, however, that the projections
obtained in that study must be regarded as a conservative estimate,
since only the contribution from the strong direct single top
production was taken into account. This is manifestly insufficient
as contributions from other channels of single top production are
indeed important, as shown in this work. Those channels will change
the kinematics of the final state partons, which might have
implications on the overall experimental efficiencies, and increase
the total cross section several times when compared with the direct
production channel alone.

The top FCNC branching ratios can also vary significantly and are
related with the total production cross sections. Although several
extensions of the SM predict significant enhancements of the
branching ratios (which in some realisations can achieve values of
the order of $10^{-4}$), most of the studies so far were done at
particle level, not taking into account the hadronization of the
final-state partons, the existence of backgrounds which might mask
the FCNC signals or the efficiencies of the detectors at the LHC.
Recent estimates from the LHC experiments~\cite{OurT7note} show
that, even when the systematic errors are very much enlarged (for
instance, by taking a conservative 5\% jet energy resolution value,
or by overestimating the contributions from the backgrounds) and
with only 1fb$^{-1}$ of data, it will be possible to gain one order
of magnitude on the current experimental limits at 95\% CL, provided
no signal of new physics is found. Assuming this very conservative
systematic error, projections can be made for other values of
luminosities and/or center-of-mass energies. If the center-of-mass
energy for the LHC is changed to only 10 TeV, the total $t\bar{t}$
production cross section changes by roughly a factor 2, which will
degrade the current estimates for the 95\% limits to
$Br(t\,\rightarrow \,q\,Z)\,\sim\, 10^{-2}$ and $Br(t\,\rightarrow
\,q\,\gamma)\,\sim\, 10^{-3}$, with only 100 pb$^{-1}$ of
luminosity. These limits are still within the interesting
experimental probing region.

In order to do a complete physics program survey, clearly a new
Monte Carlo generator must be developed which will incorporate the
contributions from all single top FCNC processes studied so far,
the hadronization of all parton level particles and the interface
with the simulation programs of the experiments if available. This
will be the subject of the next work, whereupon the predictions
stemming from cross sections calculated in this paper can be
elaborated in realistic detector scenarios.

\vspace{0.25cm} {\bf Acknowledgments:} This work is supported by
Funda\c{c}\~ao para a Ci\^encia e Tecnologia under contracts
POCI/FP/81950/2007, POCI/FP/81934/2007 and PTDC/FIS/70156/2006.
R.G.J. is supported by FCT under contract SFRH/BD/19781/2004. R.S.
is supported by the Framework Programme 7 via a Marie Curie Intra
European Fellowship, contract number PIEF-GA-2008-221707.

\appendix
\section{Feynman Rules for the FCNC vertices}

In this appendix we present the Feynman rules for the FCNC
vertices used in the calculations. In the strong sector we will
only use the Feynman rule with two quarks and one gluon. We show
the rules for a t quark entering the vertex and for a t quark
leaving the vertex.
\begin{figure}[h!]
  \begin{center}
    \epsfig{file=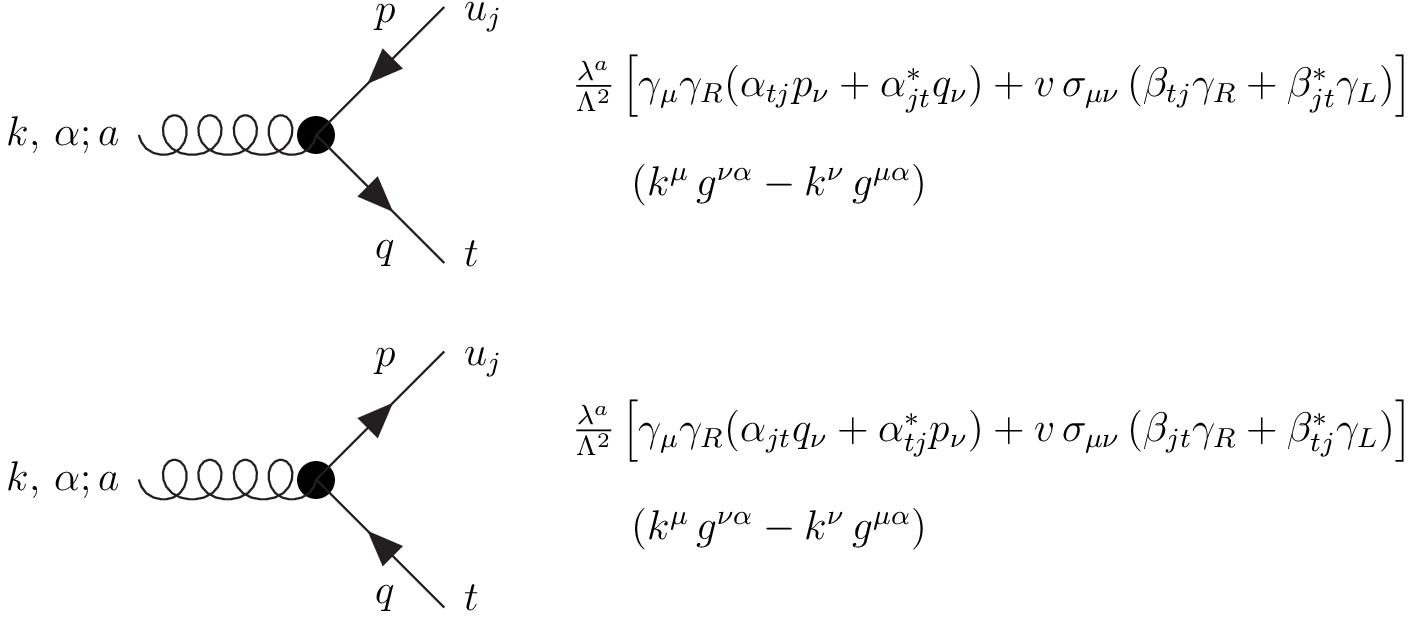,width=10.5 cm}
    \caption{Feynman rules for anomalous $g \, \bar{u_i} \, t$ and $g \, \bar{t} \, u_i$.}
    \label{fig:feynrulgluon}
  \end{center}
\end{figure}
\begin{figure}[h!]
  \begin{center}
    \epsfig{file=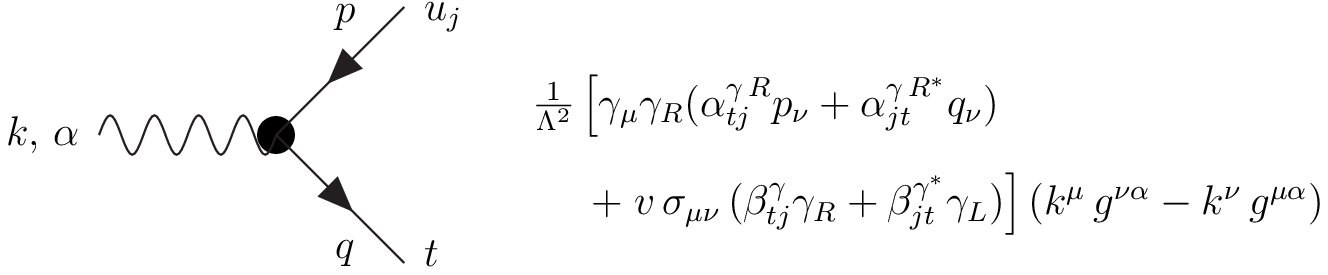,width=10 cm}
    \epsfig{file=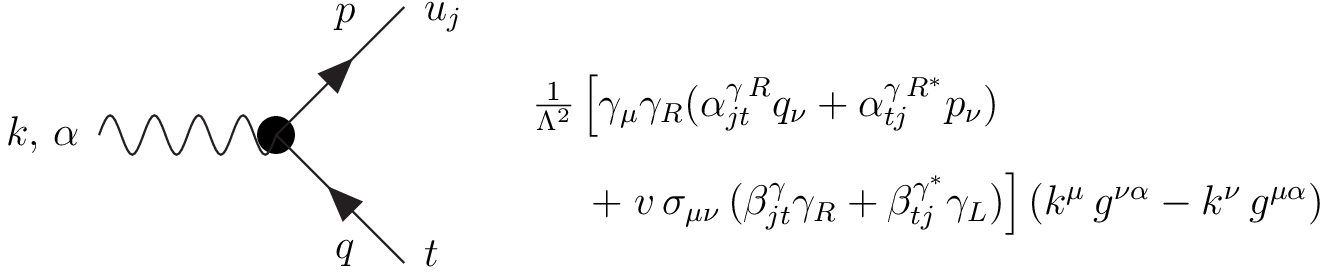,width=10 cm}
    \caption{Feynman rules for anomalous $\gamma \, \bar{u_i} \, t$ and $\gamma \, \bar{t} \, u_i$.}
    \label{fig:feyngamma}
  \end{center}
\end{figure}
\begin{figure}[h!]
  \begin{center}
    \epsfig{file=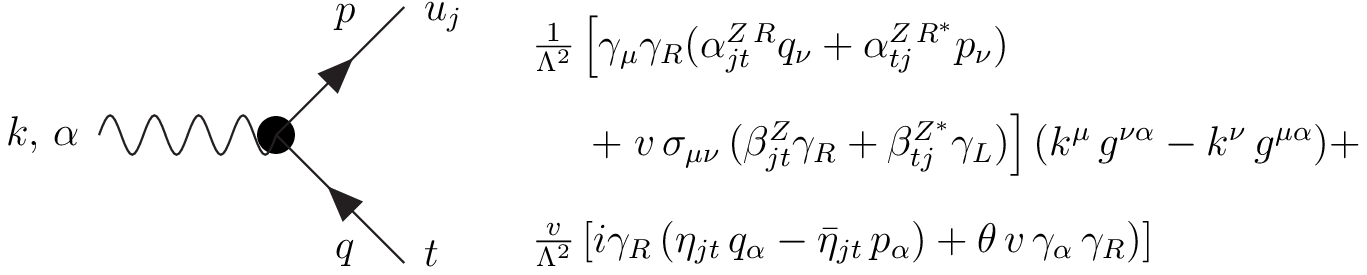,width=10 cm}
    \epsfig{file=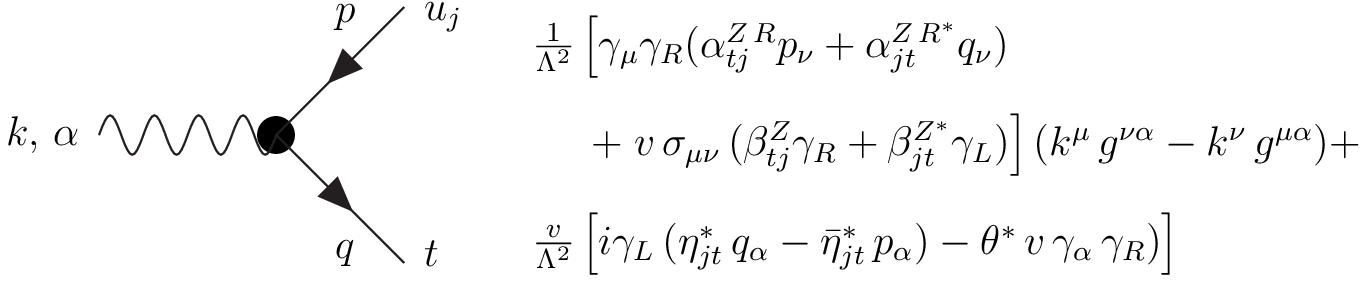,width=10 cm}
    \caption{Feynman rules for anomalous $Z \, \bar{u_i} \, t$ and $Z \, \bar{t} \, u_i$.}
    \label{fig:feynZ}
  \end{center}
\end{figure}


\begin{thebibliography}{99}

\bibitem{Ferreira:2005dr}
  P.~M.~Ferreira, O.~Oliveira and R.~Santos,
  Phys.\ Rev.\  D {\bf 73} (2006) 034011.

\bibitem{Ferreira:2006xe}
  P.~M.~Ferreira and R.~Santos,
  Phys.\ Rev.\  D {\bf 73} (2006) 054025.

\bibitem{Ferreira:2006in}
  P.~M.~Ferreira and R.~Santos,
  Phys.\ Rev.\  D {\bf 74} (2006) 014006.

\bibitem{Ferreira:2008cj}
  P.~M.~Ferreira, R.~B.~Guedes and R.~Santos, Phys.\ Rev.\  D {\bf 77} (2008)
  114008.

\bibitem{buch} W. Buchm\"uller and D. Wyler, {\em Nucl. Phys.} {\bf B268} (1986)
621.

\bibitem{Fox:2007in}
  P.~J.~Fox, Z.~Ligeti, M.~Papucci, G.~Perez and M.~D.~Schwartz,
  arXiv:0704.1482 [hep-ph].

\bibitem{toni} J.~Carvalho {\emph et al.}, {\em~Eur.~Phys.~J.}{\bf C 52} (2007)
999-1019.

\bibitem{fla} T. Lari {\em et al}, Report of Working Group 1 of the CERN Workshop
``Flavour in the era of the LHC'', hep-ph/0801.1800.

\bibitem{CMS} CMS Physics TDR: Volume II, CERN/LHCC 2006-021, http://cmsdoc.cern.ch/cms/cpt/tdr/.



\bibitem{delAguila:2000rc}
  F.~del Aguila, M.~Perez-Victoria and J.~Santiago,
  JHEP {\bf 0009} (2000) 011
  [arXiv:hep-ph/0007316].

  F.~del Aguila, M.~Perez-Victoria and J.~Santiago,
  Phys.\ Lett.\  B {\bf 492} (2000) 98
  [arXiv:hep-ph/0007160].









\bibitem{whis} E. Malkawi and T. Tait, {\em Phys. Rev.} {\bf D54} (1996) 5758;

T. Han, K. Whisnant, B.L. Young and X. Zhang, {\em Phys. Lett.}
{\bf B385} (1996) 311;

T. Han, M. Hosch, K. Whisnant, B.L. Young and X. Zhang, {\em Phys.
Rev.} {\bf D55} (1997) 7241;

K. Whisnant, J.M. Yang, B.L. Young and X. Zhang, {\em Phys. Rev.}
{\bf D56} (1997) 467;

M. Hosch, K. Whisnant and B.L. Young, {\em Phys. Rev.} {\bf D56}
(1997) 5725;

T. Han, M. Hosch, K. Whisnant, B.L. Young and X. Zhang, {\em Phys.
Rev.} {\bf D58} (1998) 073008;

K. Hikasa, K. Whisnant, J.M. Yang and B.L. Young, {\em Phys. Rev.}
{\bf D58} (1998) 114003;

T. Tait and C. P. Yuan, {\em Phys. Rev.} {\bf D63}, (2001) 014018;

D. O. Carlson, E. Malkawi, and C. P. Yuan, {\em Phys. Lett.} {\bf
B337}, (1994) 145;

T. G. Rizzo, {\em Phys. Rev.} {\bf D53}, (1996) 6218;

T. Tait and C. P. Yuan, {\em Phys. Rev.} {\bf D55}, (1997) 7300;

D. Espriu and J. Manzano, {\em Phys. Rev.} {\bf D65}, (2002)
073005.

B.~Grzadkowski, J.~F.~Gunion and P.~Krawczyk,
  Phys.\ Lett.\  B {\bf 268} (1991) 106.

\bibitem{CDFsingle}
CDF Coll., ``Combination of CDF Single Top Quark Searches with 2.2
$fb^{-1}$ of Data'', CDF Note 9251.

http://www-cdf.fnal.gov/physics/new/top/public\_singletop.html

\bibitem{D0single}
V.~M.~Abazov {\it et al.}  [D0 Collaboration], Phys.\ Rev.\  D {\bf
78} (2008) 012005.

\bibitem{CDFdirect}
CDF Coll., ``Search for top quark production via flvor-changing
neutral currents at CDF'', CDF Note 9440.

http://www-cdf.fnal.gov/physics/new/top/public\_singletop.html

\bibitem{Lept} P.~M.~Ferreira, R.~B. Guedes and R.~Santos,
{\em Phys. Rev.} {\bf D75} (2007) 055015.

\bibitem{AguilarSaavedra:2004wm}
  J.~A.~Aguilar-Saavedra,
  Acta Phys.\ Polon.\  B {\bf 35} (2004) 2695.






\bibitem{calc} M.E. Luke and M.J. Savage, {\em Phys. Lett.} {\bf B307} (1993)
387;

D. Atwood, L. Reina and A. Soni, {\em Phys. Rev.} {\bf D55} (1997)
3156;

J.M. Yang, B.L. Young and X. Zhang, {\em Phys. Rev.} {\bf D58}
(1998) 055001;

J. Guasch and J. Sol\`a, {\em Nucl. Phys.} {\bf B562} (1999) 3;

D. Delepine and S. Khalil, {\em Phys. Lett.} {\bf B599} (2004) 62;

J.J. Liu, C.S. Li, L.L. Yang and L.G. Jin, {\em Phys. Lett.} {\bf
B599} (2004) 92;

G.~Eilam, M.~Frank and I.~Turan, {\em Phys. Rev.}  {\bf D74}
(2006) 035012;

J.~J.~Cao, G.~Eilam, M.~Frank, K.~Hikasa, G.~L.~Liu, I.~Turan and
J.~M.~Yang, {\em Phys. Rev.}  {\bf D75} (2007) 075021;

A.~Arhrib and W.~S.~Hou,
 JHEP {\bf 0607} (2006) 009;

A.~Arhrib, K.~Cheung, C.~W.~Chiang and T.~C.~Yuan, {\em Phys.
Rev.} {\bf D73} (2006) 075015;

J.~Guasch, W.~Hollik, S.~Penaranda and J.~Sola, {\em
 Nucl. Phys. Proc. Suppl.}  {\bf 157} (2006) 152;

D.~Lopez-Val, J.~Guasch and J.~Sola, hep-ph/0710.0587 ;

J. A. Aguilar-Saavedra, {\em Phys. Rev.}  {\bf D67} (2003) 035003
[Erratum-ibid.\ {\bf D69} (2004) 099901];

F. del Aguila, J. A. Aguilar-Saavedra and R. Miquel, {\em Phys.
Rev. Lett.} {\bf 82} (1999) 1628;

T. P. Cheng and M. Sher, {\em Phys. Rev.}  {\bf D35} (1987) 3484;

S. Bejar, J. Guasch and J. Sola, {\em Nucl. Phys.} {\bf B600}
(2001) 21;

C. S. Li, R. J. Oakes and J. M. Yang, {\em Phys. Rev.}  {\bf D49}
(1994) 293 [Erratum-ibid. {\bf D56} (1997) 3156];

G. M.~de Divitiis, R. Petronzio and L. Silvestrini, {\em Nucl.
Phys.} {\bf B504} (1997) 45

J. L. Lopez, D. V. Nanopoulos and R. Rangarajan, {\em Phys. Rev.}
{\bf D56} (1997) 3100;

G. Eilam, A. Gemintern, T. Han, J. M. Yang and X. Zhang, {\em
Phys. Lett.} {\bf B510} (2001) 227.

\bibitem{LEP2Zgamma}
ALEPH Coll., A.~Heister  {\emph et al.}, {\em~Phys.~Lett.} {\bf
B543} (2002) 173; \\ DELPHI Coll. J.~Abdallah {\emph et al.}
{\em~Phys.~Lett.} {\bf B 590} (2004) 21; \\ OPAL Coll., G.~Abbiendi
{\emph et al.}, {\em~Phys.~Lett.} {\bf B 521} (2001) 181; \\ L3
Coll.,    P.~Achard   {\emph et al.}, {\em~Phys.~Lett.} {\bf B 549}
(2002) 290.

\bibitem{Zeus}
ZEUS Coll., S.~Chekanov {\it et al.}, {\em~Phys.~Lett.}{\bf B 559}
(2003) 153.

\bibitem{tZqCDF}
CDF Coll., ``Search for the Flavor Changing Neutral Current Decay
$t \rightarrow Zq$ in $p\bar p$ Collisions at $\sqrt{s} =
1.96$~TeV with 1.9 $fb^{-1}$ of CDF-II Data'', CDF Note 9202,
2008.


http://www-cdf.fnal.gov/physics/new/top/2008/tprop/TopFCNC\_v1.5/

\bibitem{gammaCDF} CDF Coll., F. Abe {\it et al.}, {\em~Phys.~Rev.~Lett.} {\bf 80}
(1998) 2525.

\bibitem{YR1}

  M.~Beneke {\it et al.},
  ``Top quark physics'', in "Standard Model physics (and more) at
  the LHC", G.~Altarelli and M.~L.~Mangano eds., Geneva,
  Switzerland: CERN (2000),
  [arXiv:hep-ph/0003033].

\bibitem{Ashimova:2006zc}
  A.~A.~Ashimova and S.~R.~Slabospitsky,
  arXiv:hep-ph/0604119.
H1 Coll., A. Aktas {\it et al.}, {\em~Eur.~Phys.~J.}{\bf C33},
(2004), 9.

\bibitem{gluonTevatron}
CDF Coll., V.~M.~Abrazon \textit{et al.} {\em Phys.~Rev.~Lett.}
{\bf 99}, 191802 (2007).

\bibitem{Lee}
Teh Lee Cheng, (PhD thesis), University of London, July 2007.

Sensitivity of ATLAS to FCNC single top quark production Cheng, T L;
Teixeira-Dias, P ATL-PHYS-PUB-2006-029; ATL-COM-PHYS-2006-056.-
Geneva : CERN, Aug 2006.

\bibitem{EPM}
  F.~Larios, R.~Martinez and M.~A.~Perez,
  Phys.\ Rev.\  D {\bf 72} (2005) 057504;

R.~D.~Peccei, S.~Peris and X.~Zhang,
  Nucl.\ Phys.\  B {\bf 349} (1991) 305;

T.~Han, R.~D.~Peccei and X.~Zhang,
  Nucl.\ Phys.\  B {\bf 454} (1995) 527;

R.~Martinez, M.~A.~Perez and J.~J.~Toscano,
  Phys.\ Lett.\  B {\bf 340} (1994) 91;

T.~Han, K.~Whisnant, B.~L.~Young and X.~Zhang,
  Phys.\ Rev.\  D {\bf 55} (1997) 7241.

\bibitem{miguel}
M. Won, {\em Combined effects of strong and electroweak effective
FCNC operators in top production at the LHC}, Master's Thesis,
University of Coimbra;

R.~B.~Guedes, {\em Flavour Changing at Colliders in the Effective
Theory Approach}, PhD Thesis, University of Lisbon.

Both available in
http://mars.fis.uc.pt/$\sim$plhc-top/public/publications.shtml.

\bibitem{Mertig:1990an}
  R.~Mertig, M.~Bohm and A.~Denner, Comput.\ Phys.\ Commun.\  {\bf 64} (1991) 345.

\bibitem{Bernreuther:2008ju}
  W.~Bernreuther,
  J.\ Phys.\ G {\bf 35} (2008) 083001
 and references therein;

   N.~Kidonakis,
   Phys.\ Rev.\  D {\bf 74} (2006) 114012;

   N.~Kidonakis,
 Phys.\ Rev.\  D {\bf 75} (2007) 071501.

\bibitem{Wudka:1994ny}
  J.~Wudka,
   Int.\ J.\ Mod.\ Phys.\  A {\bf 9} (1994) 2301.

\bibitem{cteq6} J. Pumplin {\em et al}, {\em JHEP} {\bf 0207} (2002) 012.

\bibitem{Cacciari:2008zb}
  M.~Cacciari, S.~Frixione, M.~M.~Mangano, P.~Nason and G.~Ridolfi,
  JHEP {\bf 0809} (2008) 127.

\bibitem{OurT7note} {\em Top quark properties}, ATLAS note, {\em in
preparation}.

\end{thebibliography}
\end{document}